# Sub-ms dynamics of the instability onset of electrospinning

**Martina Montinaro[1], Vito Fasano[1], Maria Moffa[2], Andrea Camposeo[2], Luana Persano[2], Marco Lauricella[3], Sauro Succi[3] and Dario Pisignano[1,2,*]**

[1] *Dipartimento di Matematica e Fisica "Ennio De Giorgi", Università del Salento, via Arnesano, I-73100 Lecce (Italy).*

[2] *Istituto Nanoscienze-CNR, via Arnesano, I-73100 Lecce (Italy).*

[3] *Istituto per le Applicazioni del Calcolo CNR, Via dei Taurini 19, I-00185 Rome (Italy)*

[*] *Corresponding author: dario.pisignano@unisalento.it*

**Abstract**

Electrospun polymer jets are imaged for the first time at an ultra-high rate of 10,000 frames per second, investigating the process dynamics, and the instability propagation velocity and displacement in space. The polymer concentration, applied voltage bias and needle-collector distance are systematically varied, and their influence on the instability propagation velocity and on the jet angular fluctuations analyzed. This allows us to unveil the instability formation and cycling behavior, and its exponential growth at the onset, exhibiting radial growth rates of the order of $10^3$ s$^{-1}$. Allowing the conformation and evolution of polymeric solutions to be studied in depth, high-speed imaging at sub-ms scale shows a significant potential for improving the fundamental knowledge of electrified jets, leading to obtain finely controllable bending and solution stretching in electrospinning, and consequently better designed nanofibers morphologies and structures.

## 1. Introduction

Electrospinning is a versatile, low-cost and effective technique to produce ultra-thin fibers, which has attracted a continuously increasing interest in the last decade.[1-3] The demonstrated applications of the produced nanostructures are various, including wound healing,[4,5] drug delivery,[6,7] tissue engineering,[8,9] efficient sensors,[10-12] energy harvesting,[13-15] nanophotonics[16-17] and nanoelectronics.[18-19] The advantages of electrospinning include its operational simplicity and relatively good throughput. In addition, this technique shows a good potential for tailoring the microstructure and composition of nanofibers, due to its high versatility in terms of usable compounds and blends. Consequently, electrospinning has already been demonstrated to add values to other nanofabrication approaches, such as the polymer-derived ceramic technique which has been recently used to achieve crystalline GaN nanofibers.[20]

To obtain electrospun nanofibers, a viscoelastic polymer solution is extruded from a syringe needle. The solution elongates into a collinear jet carrying an electric charge, which moves away from the needle under the effect of an externally applied electrostatic field. The collinear configuration is, however, unstable against off-axis perturbations, as one can readily realize by inspecting the effect of Coulomb repulsion on different portions of the jet. Perturbations can be in principle originated by fast micromechanical oscillations of the spinneret, and turbulences of the gas which surrounds the jet in the experimental apparatus.[21] Instead, inertial effects are usually neglected in the collinear jet, the molecular flow mostly moving in absence of internal turbulence.[2, 22] Finally, the jet reaches a collecting surface on which nanofibers are deposited in the solid state. The development of this technique has been guided by theoretical studies[23-28] and experimental works[23,29-34] concerning its fundamental characteristics. In particular, the investigation of electrospun jets and of their dynamics comes to play an important role for understanding the whole fiber fabrication process, hence it becomes the key for achieving a further and more in-depth phenomenon control. This is especially relevant for the onset of jet bending instability, which is associated to both polymer stretching and to fiber thinning during electrospinning. Jet stretching could also lead to a preferential orientation of molecular chains along the deformation direction, which in turn affects the resulting mechanical, thermal, electrical and optical properties of the fabricated nanofibers and nanocomposites.[2, 35-37] The molecular orientation or the eventual disentanglement of polymer chains can be assessed by monitoring nanofillers in X-ray experiments,[22,38] or by measuring the optical anisotropy of single electrospun nanofibers.[36] This effect would also affect interfacial interactions and phase-separation phenomena in blends or composite nanofibers.[38]

Several techniques have been developed allowing polymer solution jets to be observed,[39] thus yielding information about their conformation or evolution. Simple optical methods are undoubtedly preferable due to their immediacy, versatility in terms of imaged fields, and ease of integration with electrospinning equipment. However, a few artifacts can be present in the resulting jet micrographs, such as glints due to jet-reflected light or interference colors.[39] Most of reported optical techniques to investigate electrospinning are based on multiple and quite complex imaging components and illumination sources. Stereographic images of electrospun jets have been collected, using prisms during light collection and then viewing stereoscopically so-obtained side-by-side pictures.[23] The resulting micrographs provide meaningful information on jet velocity and on its three-dimensional displacement, although the jet position is averaged owing to the quite long imaging times, of order of 16-30 ms.[23,39] Furthermore, velocity measurements of electrospun jets have been performed employing laser Doppler velocimetry[40,41] which is based on the employment of coherent laser beams and permits, in principle, to measure jet velocity components in any direction.[39]

Another method largely exploited in order to facilitate the jet visualization and analysis is the insertion of small particles in electrospun polymer solutions.[39] For instance, silica microbeads can be traced with an X-ray beam to gain information about the flow regime in the initial, straight part of the jets.[22] Light-emitting particles can also be tracked, allowing small transverse oscillatory movements, probably related to the instability onset in the jet, to be captured by fluorescence microscopy, with overall temporal resolution in the range of ms.[42] In fact, particles serve as probes which increase significantly imaging contrast with respect to purely polymeric solutions, hence the actually measured velocity is mainly related to particles themselves. This velocity can be associated to the jet, supposing that particles consistently follow the flow of the carrying polymer solution, and that the jet properties and behavior are not remarkably changed following particle introduction. Moreover, the related solution preparation and experimental setup are possibly complex, needing to avoid issues of particles aggregation and eventual needle clogging, and relatively small regions (< mm) of the jet are typically imaged effectively, nearby the extruding tip. Finally, the need of small fluorescent particles (ideally sub-µm), suitable for probing thin jets, limits collectable intensity signals.

In this framework, optical imaging at very high frame-rates can lead to a more in-depth understanding of the electrospun jet phenomenology related to the instability onset. Surprisingly given the wide success of electrospinning technologies, there is a lack of data about the real-time evolution of the path of polymer solution jets, due to the generally limited temporal resolution of used characterization methods, which only partially match the characteristic times involved in the process dynamics. The present work reports on electrospun polymer jets investigated through imaging with a rate of 10,000 frames per second (fps). This allows the sub-ms timescale of the jet dynamics to be unveiled and analyzed, namely the onset of electrospinning bending instability to be appreciated in real time. The developed set-up is easy to implement, thus allowing researchers to observe electrified jets at high frame-rates with no need for complex experimental facilities. We investigate the instability propagation velocity and the instantaneous angular aperture of jets under different conditions of polymer solution concentration, needle-collector distance and applied voltage. The characteristic stages of the instability evolution are so observed and their cyclic repetition at high frequency outlined. In particular, managing times comparable to those of the involved multi-physics process, we observe the exponential growth of off-axis displacements as sketched in early works by Reneker and coworkers.[23] This growth is found to occur at rates of the order of $10^3$ $s^{-1}$ in its initial stage, highlighting the sub-ms character of the instability onset dynamics.

## 2. Materials and methods

### Electrospinning process

As prototypical investigated system, poly(vinylpyrrolidone) (PVP, molecular weight = 1300 kDa, Alfa Aesar) is used. The polymer is dissolved in a mixture of ethanol/water (17:3 v:v), at a concentration ranging between 11 and 21 mg/mL. Lower concentrations do not lead to reliable electrospinning processes, whereas higher concentrations caused frequent needle clogging and discontinuous spinning operation. Electrospinning is performed by placing the solution into a syringe tipped with a 21-gauge stainless steel needle and connected to a high voltage supply (XRM30P, 82 Gamma High Voltage Research) for the application of bias values in the range 6-11 kV. A 33 Dual Syringe Pump (Harvard Apparatus Inc., Holliston, MA) is used to supply a constant flow rate of 2 mL/h. All processes are carried out in air, with temperature and humidity values of 21-24 °C and 30-35%, respectively. A static collector consisting of a metallic plate is placed at distance varying between 14 cm and 26 cm from the needle.

### Polymer jet imaging

The system used for imaging the polymer jet is constituted by a high speed camera (FAST CAM

APX RS Photron, San Diego, California) allowing for collecting videos (512 pixel×512 pixel) at 10,000 fps, corresponding to acquisition times of 0.1 ms. In such configuration, the maximum temporal duration of each continuously acquired video is 800 ms. The polymer jet is imaged by using different objectives, either with variable focal length (in the range 18-55 mm, Nikkor, Japan) or with fixed focal length (200 mm, Nikkor, Japan), having $f$ number 3.5-5.6 and 4 respectively. The camera is installed on a translation stage, allowing the position with respect to the polymer jet to be precisely controlled in the vertical and horizontal directions. The polymer jet is illuminated by a halogen lamp, placed at a distance of 60 cm from the needle. A semi-transparent screen is interposed between the lamp and the electrospinning region, to diffuse light and to prevent significant heat from the lamp reaching the jet. Temperature changes of about 1 °C are measured by a thermocouple in the jet region upon varying the illumination conditions. No glints are observed during experiments. All the images used for the analysis are corrected by subtracting a background image.

## 3. Results and Discussion

Videos collected with temporal resolution of 0.1 ms allow the whole dynamics of an electrified viscoelastic jet nearby the onset of the instability regime to be appreciated. For our PVP solutions, the jet propagates in a straight line for a few centimeters after extrusion from the syringe orifice. Then, the first electrical bending instability takes place. Interestingly, focusing the camera on the straight segment we can observe a sequence of repeated phases characteristic of the ongoing physical process. Fig. 1 presents a series of frames depicting the typical behavior of a PVP solution of concentration of 20 mg/mL, spun with an electric field of 69 kV/m. The here shown frames refer to a region which is located at about 4 cm from the needle, and illustrate the entire instability sequence, from its formation and propagation to its vanishing, after which a new fairly identical sequence begins. The characteristic times mentioned in the following are average values obtained from a statistics over at least ten cycles. Five consecutive stages can be considered as follows. In particular, the high temporal resolution allows us to observe in detail the sub-phases (steps 1-3 listed below), corresponding to the transition of a straight segment of jet to a slightly curved structure, which then leads to the formation of a three-dimensional spiral.[43]

1. Following extrusion, the jet extends along a straight line over a few cm. We hereby arbitrarily fix as time zero ($t$ =0 ms in Fig. 1) the initial observation instant, at which the jet still exhibits a well-defined linear shape.

2. Then a small perturbation from the linear trajectory is developed, from which a bending instability arises ($t$=3.5 ms in Fig. 1). This onset depends on specific jet parameters and conditions,[44] as discussed below.

3. The perturbation widens and the jet generates an expanding coil also in the opposite side of the first wave peak, along a direction roughly perpendicular to the jet longitudinal axis and within the imaged plane (namely, the focal plane of the imaging optical system, identified by the plane normal to the optical axis and to which the jet straight segment belongs). Then, for about 6 ms the jet undergoes a series of bending coils that propagate along its longitudinal axis: to sustain the oscillation caused by a broadening coil, the jet bends again producing a concomitant folding coil that enlarges in the opposite direction with respect to straight jet axis. Examples of this quasi-bidimensional propagation are represented in frames at times between 5.3 ms and 11.6 ms in Fig. 1.

4. Then, for about 18 ms the jet develops a tridimensional motion reaching regions out from the focal plane, at least at a distance of 300 μm in whichever direction. The jet blurry inferior extremity, clearly visible in frames relative to times between 13.3 and 20.0 ms in Fig. 1, evidences that the jet is already looping and twisting around its own longitudinal axis, beyond the depth of focus of the acquisition system.

5. After the stage of tridimensional propagation, the jet closes up and returns on the focal plane, only slightly whipping and oscillating nearby its own axis. These stabilization movements last for about 9 ms, at the end of which the jet path recovers its nearly straight line shape ($t = 24.3$ ms in Fig. 1). In this condition, the jet is ready to experience a new small initial perturbation that will provoke a new bending instability, beginning again this cycle.

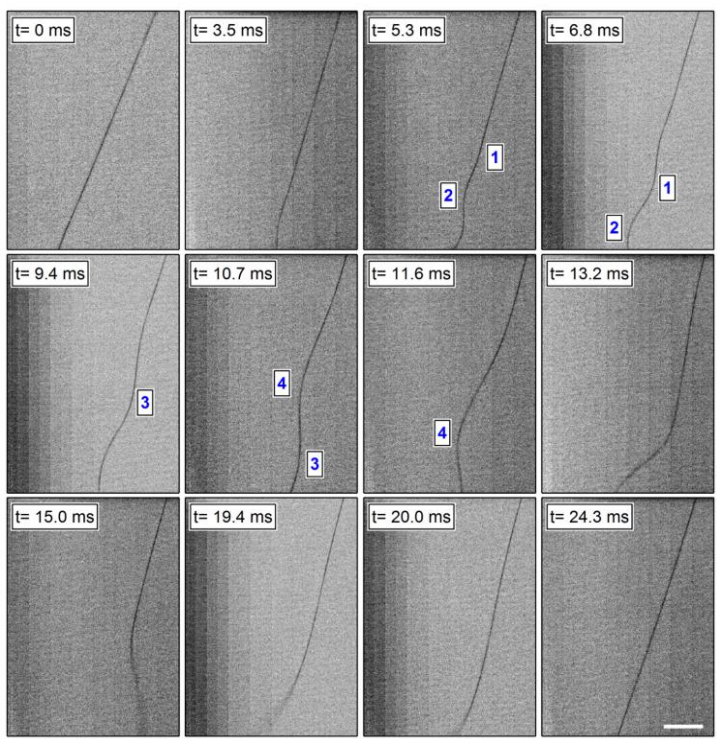

**Fig. 1**. A sequence of frames of a typical instability cycle of a 20 mg/mL PVP solution ($d = 16$ cm, $V = 11$ kV). Time zero (first frame) is taken in correspondence to the occurrence of the initial straight shaped jet. Progressive numbers in the micrographs highlight the evolution of each individual coil over time. Scale bar = 2 mm.

Bending instability establishes with a frequency of about 30 Hz, each whole cycle lasting for about 35 ms. In particular, we clearly observe that the first bending propagates initially downward and, only at a second stage, also outward the initial jet axis direction. After broadening into a three-dimensional path, the jet assumes again a linear trajectory along its longitudinal axis, until a new bending instability arises. In turn, the three-dimensional perturbation will elongate and establish ample spirals as detailed in Ref. 43. The growth of these spirals on a larger scale and the formation of eventual, secondary oscillating instabilities take place on a spatial and temporal window external to the frames shown in Fig. 1. The dynamics here observed also significantly differs from previously observed whipping and pendulum-like motions,[45] which occur in the initial and straight region of jets well before the establishment of bending instability.

**Velocity measurements and jet path**

Following the propagation of single coils within an optical field of view, it is possible to measure their velocity with a very low relative error (about 0.4%), extracting the spatial positions of the growing and bending instability maxima at each frame and exploiting the concomitantly high temporal and spatial resolutions of the acquisition apparatus. An example of such analysis for a well-definite coil peak is shown in Fig. 2, where an instability peak is highlighted with a bright dot in each frame. The right-bottom panel in Fig. 2 summarizes all the positions taken by the peak during its propagation along the jet axis over 7 ms. The resulting tracking arrows are indicative of velocity, and are characteristic of the solution and of the process parameters.

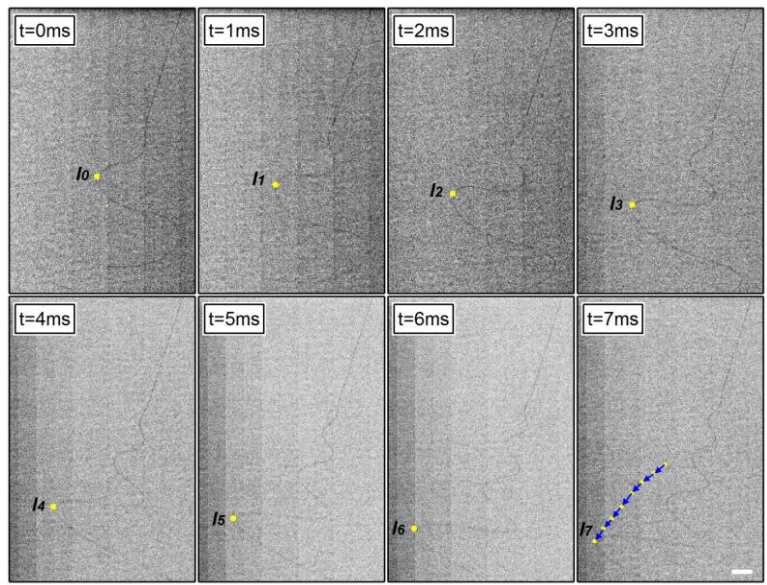

**Fig. 2.** Sequence of frames for determining instability velocity. The motion of the peak under observation is highlighted by bright dots, indicating the consecutive positions of the coil apex during its propagation at step-intervals of 1 ms. A sketch of the resulting velocity vector analysis is superimposed to the jet micrograph in the last panel ($t$ = 7 ms). Scale bar = 2 mm.

Fig. 3a compares averaged peak positional data relative to electrospinning processes at different needle-collector distances $d$ (keeping constant all other parameters). The linear fits of the positions ($l$) provide velocity values. The jet average velocity $v_m$ is found to increase for decreasing needle-collector distance, due to the correspondingly increasing electric field, as shown in Fig. 3b.

Overall, the studied instability peaks propagate with a velocity of 1.5-4.0 m/s along the jet. We also perform a parametric study about both the average velocity (Fig. 4) and the instantaneous angular aperture (Θ) of the instability cone as visualized in the focal plane (Fig. 5). In the range of applied voltages between 6 and 11 kV, we find a monotonous increase of the measured velocity for increasing voltage values, again consistent with a concomitant increase of the electric field (Fig. 4a). Differently, the solution concentration is not found to affect velocity significantly (Fig. 4b).

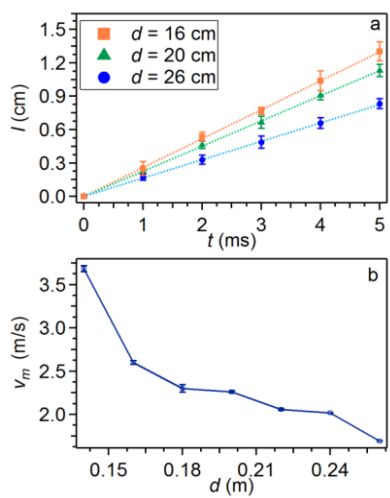

**Fig. 3.** (a) Experimental study of jet velocity. 20 mg/mL PVP solution processed at $V = 9$ kV for different values of $d$. The needle-collector distances of 16 cm, 20 cm and 26 cm correspond to an average velocity value of 2.6 m/s, 2.2 m/s and 1.6 m/s respectively, with relative errors of 0.4%. (b) Instability average velocity $v_m$ of the jet vs. $d$.

As sketched in Fig. 5a, $\Theta$ angles relate to jet spiraling in a direction perpendicular to its initial longitudinal axis and can be easily inferred by individual frame micrographs. This radial displacement is counteracted by the polymer viscoelastic resistance. Indeed, while a weak or no effect is found on $\Theta$ upon varying the applied electric field through either $d$ (Fig. 5b) or $V$ (Fig. 5c), the viscoelastic response is found to be strongly enhanced by more concentrated jet solutions through a more pronounced fluid stiffness, as found experimentally and shown in Fig. 5d. Measuring $\Theta$ continuously over time, we obtain fluctuation curves as those shown in Fig. 6. The insets in the graphs show corresponding Fast Fourier Transform (FFT) curves, evidencing a white noise behavior for high frequency ($f$~a few kHz), preceded by an $f^{-1}$- $f^{-2}$ behavior at lower frequencies. The specific fluctuations observed largely vary from one process to another, thus being likely related to vibrations or other noise sources in the spinning environment.

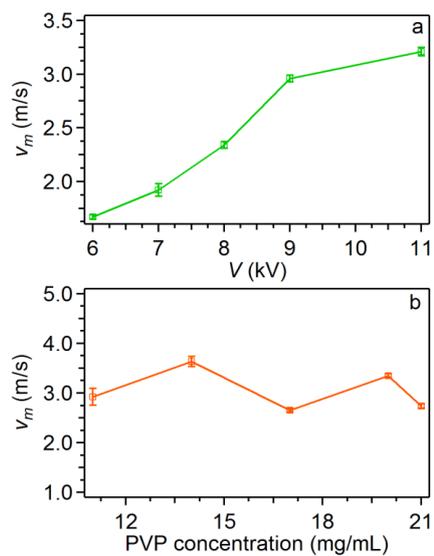

**Fig. 4.** Instability average velocity vs. applied voltage (a) and polymer solution concentration (b). Other parameters are: PVP concentration = 20 mg/mL (a), $d$ = 16 cm (a, b), $V$ = 9 kV (b).

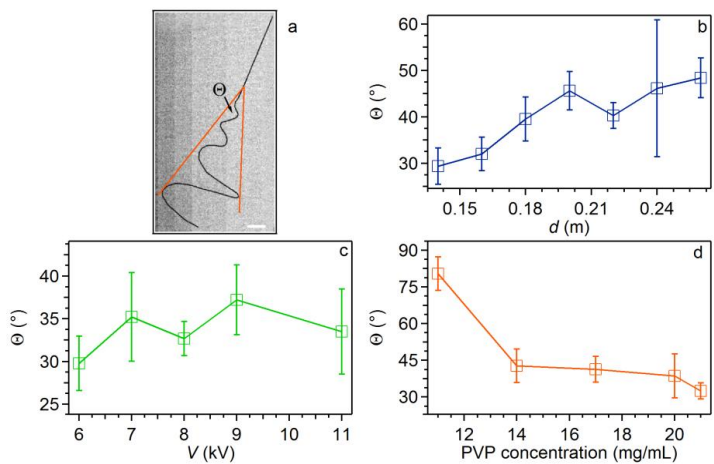

**Fig. 5.** (a) Micrograph showing the instability broadening and highlighting the resulting angular aperture (Θ). The jet is manually contrasted for better clarity. Scale bar = 2 mm. (b-d) Θ values vs. needle-collector distance (b), applied voltage (c) and polymer solution concentration (d). Other parameters are PVP concentration = 20 mg/mL (b, c), $d$ = 16 cm (c, d), $V$ = 9 kV (b, d).

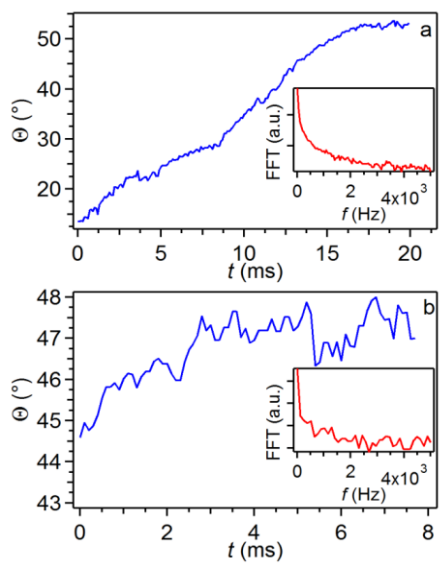

**Fig. 6.** Examples of fluctuations of the angle Θ. Curves here denote a constant growth in the jet angular aperture in the investigated temporal windows. Insets: corresponding FFT curves. PVP solution concentration = 20 mg/mL, $d$ = 24 cm, $V$ = 9 kV.

**Electrical bending instability onset**

The length of the initial straight segment, $h$, is between about 0.5 cm and 7 cm, and it is found to increase upon increasing the applied voltage bias and for more concentrated solutions, whereas it decreases for higher needle-collector distances (Fig. 7). These findings are in agreement with previous results on electrospun polyethylene oxide solutions.[43] These results confirm that the straight segment stability is strictly related to the longitudinal stress applied by the electric field.[39,41] On the other hand, also the polymer concentration plays an important role in this respect, since denser solutions counteract the establishment of radial movements which are precursors of bending instabilities. Indeed, we find that $h$ values correlate with Θ angles, higher $h$ values being accompanied by generally lower angular aperture of the jet, as shown in Fig. 8. The general trend of increasing values of $h$ for decreasing values of $\Theta$ is kept for the variation of whichever parameter ($d$, $V$ and polymer concentration), $h$ and $\Theta$ relating therefore both to underlying electrical instabilities. Furthermore, Fig. 8 leads to immediate conclusions concerning the investigated parameters space, highlighting the polymer solution concentration as parameter which mostly affects the behavior of the electrified jet. Analogously, variations of the needle-collector distance influence the jet behavior more relevantly than the applied voltage. For a more detailed analysis, one should also consider the role of the polymer chemistry, as well as the interplay between the onset of the bending instability and polymer entanglements. The specific combination of the degree of networking of macromolecules in the solution, steric hindrance and relaxation time of a given polymer species may lead to a specific jet behavior under electrospinning strain. For instance, low entanglements under high strain rate extensional flows may lead to jet breaking and short nanofibers.[46] The overall solution viscoelasticity, together with the charge density carried by the jet, and the surface tension are demonstrated to be key parameters for the formation of beaded electrospun nanofibers.[43,47] Local charge imbalance and defect density within polymeric bonds may also affect the resulting configuration of stretched macromolecules.[48]

   To a first approximation, the stability of electrospun jets is determined by the balance between stabilizing surface tension and electrical repulsion between adjacent charges along the jet.[24] In particular, following a small initial perturbation, Coulomb repulsion forces between charges can easily cause the system to undergo bending and movement in a radial direction as illustrated by Reneker and coworkers.[23] In fact, the initial perturbation, when comparable to jet diameter, would undergo a rapid, exponential growth in a direction perpendicular to the initial straight segment, thus generating a coil.[39] In other words, this growth is driven by the decreasing potential electrostatic energy associated to radial displacement. The coil growth would roughly follows a law:

$$\Delta = \Delta_0 \exp(\sigma t), \tag{1}$$

where $\Delta$ is the distance of the coil peak from the jet axis (inset of Fig. 9), $\Delta_0$ is the initial small radial perturbation, $\sigma$ is a characteristic rate constant depending on the system properties and $t$ is time.[23] The $\Delta$ distance is in turn reflected in an angular displacement ($\alpha$) with respect to the initial longitudinal axis of the jet [$\alpha = \mathrm{tg}^{-1}(\Delta/\lambda)$, where $\lambda$ indicates an instability wavelength-scale length along the jet axis, as schematized in the inset of Fig. 9]. For small displacement, i.e. at very short times, the previous relation leads to:

$$\alpha = \alpha_0 \exp(\sigma t), \tag{2}$$

with $\alpha_0$ angular displacement at $t \ll 1$. Times involved in the radial growth process are of the order of 0.1-1 ms (corresponding to $\sigma$ rates as high as $10^3$-$10^4$ s$^{-1}$). Here, measuring angles with sub-ms temporal resolution and with spatial resolution of about 20 µm/pixel, we can therefore clearly observe the exponential growth of the first bending instability as predicted by Eq. (2). For instance, for a PVP solution with a concentration of 20 mg/mL, $V$=9 kV and $d$=16 cm, the $\sigma$ rate is of 760 s$^{-1}$ from fitting experimental data, corresponding to a characteristic time of about 1.3 ms (Fig. 9). Interestingly, such rate agrees well with typically reported strain rates ($10^3$ s$^{-1}$) of jets during bending loops.[23] Moreover, this value compares well with the strain rate resulting from considering the cross-sectional area reduction ratio of the electrified jet.[43] Indeed, for an initial jet diameter of 0.5 mm and a final diameter of a dry, deposited fiber of about 3 µm, one obtains an overall area reduction ratio of the order of $10^4$. Taking into account the solvent evaporation, this value reduces to about 450, which is the area reduction ratio due to the overall elongational strain along the jet. PVP solutions would suffer a corresponding draw ratio of the same order of magnitude. This estimation, with a time of flight of the jet of a few tenths of ms, would lead to a longitudinal strain rate of about $10^3$ s$^{-1}$.[43]

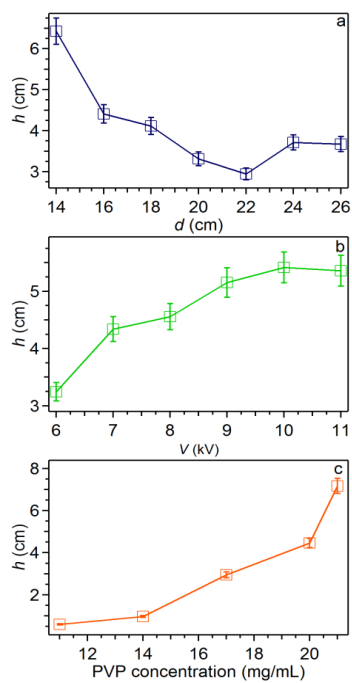

**Fig. 7.** $h$ values vs. needle-collector distance (a), applied voltage (b) and polymer solution concentration (c). Other parameters are PVP concentration 20 mg/mL (a, b), $d = 16$ cm (b, c), $V = 9$ kV (a, c).

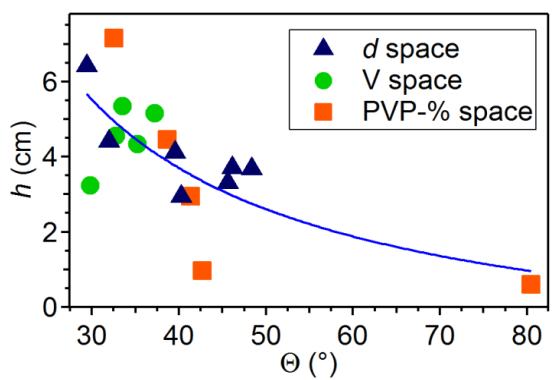

**Fig. 8.** Plot of $h$ vs. average $\Theta$ values. Triangles: data collected in the distance series ($d = 14$-$26$ cm), circles: voltage series ($V = 6$-$11$ kV), squares: PVP concentration series (11-21 mg/mL). The continuous line is a guide for the eye.

Overall, this study supports the richness of phenomenology and interplay of various process parameters characterizing electrospinning, providing first quantitative data about the very initial stage of bending instability. In order to analyze the instability character at the sub-ms onset more in depth, and relate its growth rates to polymer solution properties, fast imaging could be applied to a large number of macromolecule systems, and to their solutions with different degrees of entanglement. This would enable a more direct understanding of molecular factors favoring or depressing bending instabilities, which are in turn directly related to the morphology, diameter and architectures of ultimately deposited nanofibers. Few subjects of future research are naturally connected with the present experimental investigation. First, it would be of interest to study the effects of coherent hydrodynamic motion on the jet dynamics, as well as on the onset of the bending instability. Indeed, given that the system could be able to support coherent structures in the wake of the propagating jet, a slight asymmetry between such coherent structures may affect, or maybe even trigger, the bending instability. Another subject of interest would be the development of some form of statistical mechanics of the jet propagation. Typically, one could ask what is the probability $P(z, w; t)$ of finding a fluid body ejected at the nozzle ($z = 0$ at time $t = 0$), at elevation $z$ at time $t$ with velocity $w$. In fact, this procedure is tantamount to looking for the Green function, or propagator, of the kinetic equation governing the statistics of an ensemble of trajectories, each generated, for instance, by different initial conditions, say noise in the spinneret or in the environment. A natural candidate for this task is the following Fokker-Planck equation:

$$\partial_t P + w \partial_z P = \partial_w \left[ (F(z) - \gamma w) P + D\, \partial_w P \right] \qquad (3)$$

with initial conditions $P(z, t = 0) = \delta(z)$. The above equation associates with an effective Langevin equation, $\dot{z} = w$ and $\dot{w} = F(z)/m - \gamma w + \zeta$, where $\zeta$ is a stochastic noise obeying the fluctuation-dissipation theorem $<\zeta(t)\zeta(t')> = D\delta(t-t')$, with $D = k_B T/m$ being the diffusion coefficient.[49] Manifestly, the above

Fokker-Planck equation is a drastic simplification of the actual process, as it incorporates all the many-body conservative interactions within an effective one-body force $F(z)$. In addition, the full three-dimensional structure of the jet should be tracked in time, not just the $z$ coordinate. Nevertheless, Eq. (3) is most likely to provide interesting new insights into the statistical dynamics of the jet. In passing, we note that a very similar approach has indeed proven very useful for the case of DNA translocation through nanopores.[50,51] In fact, it is quite possible that much could be learned about the statistical mechanics of electrospinning by drawing a judicious parallel with biopolymer translocation through nanopores and biological membranes.

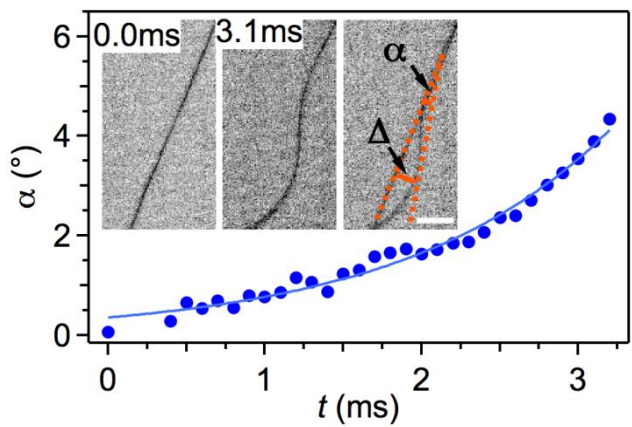

**Fig. 9.** Exponential growth of the angular displacement, α, at the instability onset for an electrospun PVP solution jet (solution concentration = 20 mg/mL, $V$ = 9 kV, $d$ = 16 cm). The solid line is a fit to experimental data by Eq. (2). Insets: α variation within an interval of about 3 ms. The rightmost inset is the overlap of the previous two frames, highlighting the resulting α angle. Scale bar = 0.5 mm.

## 4. Conclusions

Solutions with PVP concentrations of 11-21 mg/mL have been electrospun, and jets observed by fast optical imaging at 10,000 fps. This analysis provides information on the characteristic times involved in the process, on the instability formation, on average velocities (1.5-4.0 m/s in this study) and on the instantaneous jet angular aperture and its dependence on process parameters. Furthermore, the exponential growth of the instability at its initial stages has been experimentally observed, and found to occur at rates of the order of $10^3$ $s^{-1}$. How this growth rate can vary with process parameters for various polymer species is an interesting topic of forthcoming research, aiming at obtaining finely controllable jet bending and solution stretching, and consequently better designed nanofibers diameter and morphology.

Providing direct information on polymeric solution jets and on their rapid evolution, sub-ms optical imaging may therefore significantly contribute in improving both the knowledge of fundamental aspects of electrospinning science and the capability to investigate the process space of parameters and their influence on ultimately resulting nanofibers. This kind of observation is not limited to the initial, linear section of electrified jets. For these reasons, it might stimulate new research and the development of ultra-fast characterization methods applying to electrospinning.

**Acknowledgements**

The research leading to these results has received funding from the European Research Council under the European Union's Seventh Framework Programme (FP/2007-2013)/ERC Grant Agreements n. 306357 (ERC Starting Grant "NANO-JETS").